\documentclass[twoside,a4paper,11pt]{sea10}
\usepackage{graphicx}
\usepackage{hyperref}
\usepackage{movie15}
\usepackage{pdflscape}
\def\farcs{\hbox{$.\!\!^{\prime\prime}$}}
\def\degr{\hbox{$^\circ$}}
\def\nodata{ ~$\cdots$~ }
\begin{document}
\pagenumbering{arabic}
\pagestyle{myheadings}
\thispagestyle{empty}
{\flushleft\includegraphics[width=\textwidth,bb=58 650 590 680]{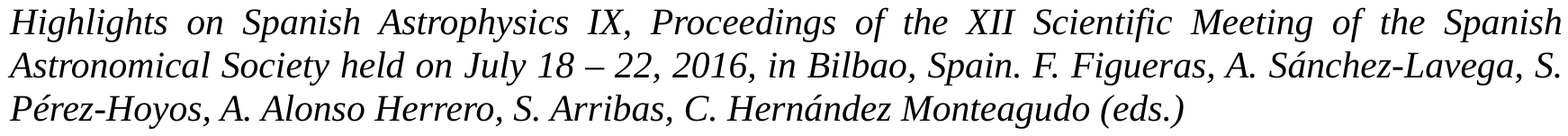}}
\vspace*{0.2cm}
\begin{flushleft}
{\bf {\LARGE
%
The Galactic O-Star Catalog (GOSC) and the Galactic O-Star Spectroscopic Survey (GOSSS): current status
%
}\\
\vspace*{1cm}
%
Jes\'us Ma{\'\i}z Apell\'aniz$^{1}$,
\'Alvaro Alonso Morag\'on$^{2}$, 
Luc{\'\i}a Ortiz de Z\'arate Alcarazo$^{2}$, 
and 
the GOSSS team
%
}\\
\vspace*{0.5cm}
%
$^{1}$
Centro de Astrobiolog\'{\i}a, CSIC-INTA, Madrid, Spain\\
$^{2}$
Universidad Complutense, Madrid, Spain
%
\end{flushleft}
%
\markboth{
GOSC and GOSSS: current status
}{ 
%
Ma{\'\i}z Apell\'aniz et al.
%
}
\thispagestyle{empty}
\vspace*{0.4cm}
\begin{minipage}[l]{0.09\textwidth}
\ 
\end{minipage}
\begin{minipage}[r]{0.9\textwidth}
\vspace{1cm}
\section*{Abstract}{\small
%
We present the updates of the Galactic O-Star Catalog (GOSC) that we have undertaken in the last two years: new spectral
types, more objects, additional information, and coordination with CDS. We also present updates for the Galactic O-Star 
Spectroscopic Survey (GOSSS). A new paper (GOSSS-III) has been published and $\sim$1000 targets have been observed since 2014.
Four new setups have been added to our lineup and for two of them we have already obtained over 100 spectra: with OSIRIS at the 
10.4~m GTC we are observing northern dim stars and with
FRODOspec at the 2.0~m Liverpool Telescope we are observing northern bright stars. Finally, we also make available new versions of 
MGB, the spectral classification tool associated with the project, and of the GOSSS grid of spectroscopic standards.
%
\normalsize}
\end{minipage}
%
%
%
\section{What are GOSC and GOSSS?}

$\,\!$\indent GOSC, the Galactic O-Star Catalog, is an on-line catalog for O stars (and relatives): \url{http://gosc.iaa.es}.
The first version was presented by \href{http://adsabs.harvard.edu/abs/2004ApJS..151..103M}{Ma{\'\i}z Apell\'aniz et al. (2004)}
and the first major update by \href{http://adsabs.harvard.edu/abs/2008RMxAC..33...56S}{Sota et al. (2008)}. The current
version is v3.2.2 an it has two sections: one public with 601 objects and another private with 7000+ objects.

GOSSS, the Galactic O-Star Spectroscopic Survey (\href{http://adsabs.harvard.edu/abs/2011hsa6.conf..467M}{Ma{\'\i}z Apell\'aniz et al. 2011}), 
is observing all optically accessible Galactic stars that anybody has ever classified as O (if we get time on a large enough telescope).
It obtains $R\sim$ 2500 spectroscopy in the blue-violet region with S/N $\sim$ 250. It has currently observed
2500+ stars, including all O stars down to complete to $B=8$, with some objects as dim as $B=17$.
In some cases, mostly for extreme SB2s and variables, we have multiple epochs. As of 2016 we have published three major blocks of 
the survey: GOSSS-I (\href{http://adsabs.harvard.edu/abs/2011ApJS..193...24S}{Sota et al. 2011}), GOSSS-II 
(\href{http://adsabs.harvard.edu/abs/2014ApJS..211...10S}{Sota et al. 2014}), and GOSSS-III 
(\href{http://adsabs.harvard.edu/abs/2016ApJS..224....4M}{Ma{\'\i}z Apell\'aniz et al. 2016}).

\section{GOSC+GOSSS goals}

$\,\!$\indent The primary goal of GOSC and GOSSS is the spectral classification of O stars. More specifically we aim to [a]
identify and classify all optically accessible Galactic O stars, [b] improve classification criteria and possibly define new 
special types and [c] identify objects wrongly classified as O. They also have five secondary goals:

\begin{enumerate}
 \item Derive the physical properties of O stars such as effective temperature or $v\sin i$.
 \item Study SB2s in collaboration with their high-resolution sister surveys (OWN, CAF\'E-BEANS, IACOB, and NoMaDS).
 \item Improve our knowledge of the extinction law and the ISM.
 \item Measure the spatial distribution of massive stars and dust.
 \item Obtain the massive-star IMF.
\end{enumerate}

\section{What is new in GOSC in the last two years?}

$\,\!$\indent In the last two years GOSC has produced several new versions, going from from v3.1.1 to v3.2.2. The most important
update has been the use of GOSSS DR2.0 from GOSSS-III 
(\href{http://adsabs.harvard.edu/abs/2016ApJS..224....4M}{Ma{\'\i}z Apell\'aniz et al. 2016}) for public spectral types (those in
the main catalog and supplements 1-4, see Table~1). Furthermore, the following updates have also been included:

\begin{itemize}
 \item There are 1500 new objects, including all bright B stars. 
 \item We have added WISE H\,{\sc ii} regions in the clusters and nebulae field.
 \item Coordinates have been revised with new Tycho-2 and 2MASS data.
 \item We have added 2MASS catalog numbers and Simbad names.
 \item We have added identifiers (CPD, BD, and ALS).
 \item We are coordinating with the CDS team at Strasbourg for updates on Simbad spectral types.
\end{itemize}

\begin{table}
\caption{Main catalog and supplements in GOSC.}
\center
\begin{tabular}{ccl}
\multicolumn{1}{l}{ Name} & \multicolumn{1}{l}{ Public?} & Description \\
\hline
Main & Yes & Galactic O stars \\
S01  & Yes & Galactic WR and WR+O systems \\
S02  & Yes & Other Galactic early-type stars \\
S03  & Yes & Galactic late-type stars \\
S04  & Yes & Extragalactic massive stars \\
S05  & No  & Galactic O stars \\
S06  & No  & Galactic WR and WR+O systems \\
S07  & No  & Other Galactic early-type stars \\
S08  & No  & Galactic late-type stars \\
S09  & No  & Extragalactic massive stars \\
S10  & No  & Galactic early-type candidates not observed with GOSSS \\
S11  & No  & Galactic late-type candidates not observed with GOSSS \\
S12  & No  & Extragalactic candidates not observed with GOSSS \\
S13  & No  & Close companions unresolved in GOSSS \\
\hline
\end{tabular}
\label{table1}
\end{table}

\section{What is new in GOSSS in the last two years?}

$\,\!$\indent In the last two years the most important addition to GOSSS has been the publication of GOSSS-III 
(\href{http://adsabs.harvard.edu/abs/2016ApJS..224....4M}{Ma{\'\i}z Apell\'aniz et al. 2016}), see section 5.
We have also produced new versions of MGB (section 6) and of the OB2500 standard grid (section 7).

We have also observed $\sim$1000 objects in the last two years alone and we have added four new telescopes to our lineup:
GTC, Liverpool, SOAR, and Gemini South. Of those, we emphasize GTC and Liverpool, as they have become two of our workhorses,
observing hundreds of objects with each one of them. At the 10.4~m GTC we use OSIRIS to observe northern dim stars 
($14 < B < 17$) with filler programs. At the robotic 2.0~m Liverpool we are using FRODOspec, a miniIFU spectrograph, to observe
northern bright stars ($11 < B$). For the latter case we have developed a specific pipeline, as all other setups use long slits.

\begin{figure}
\centerline{\includegraphics*[width=\linewidth]{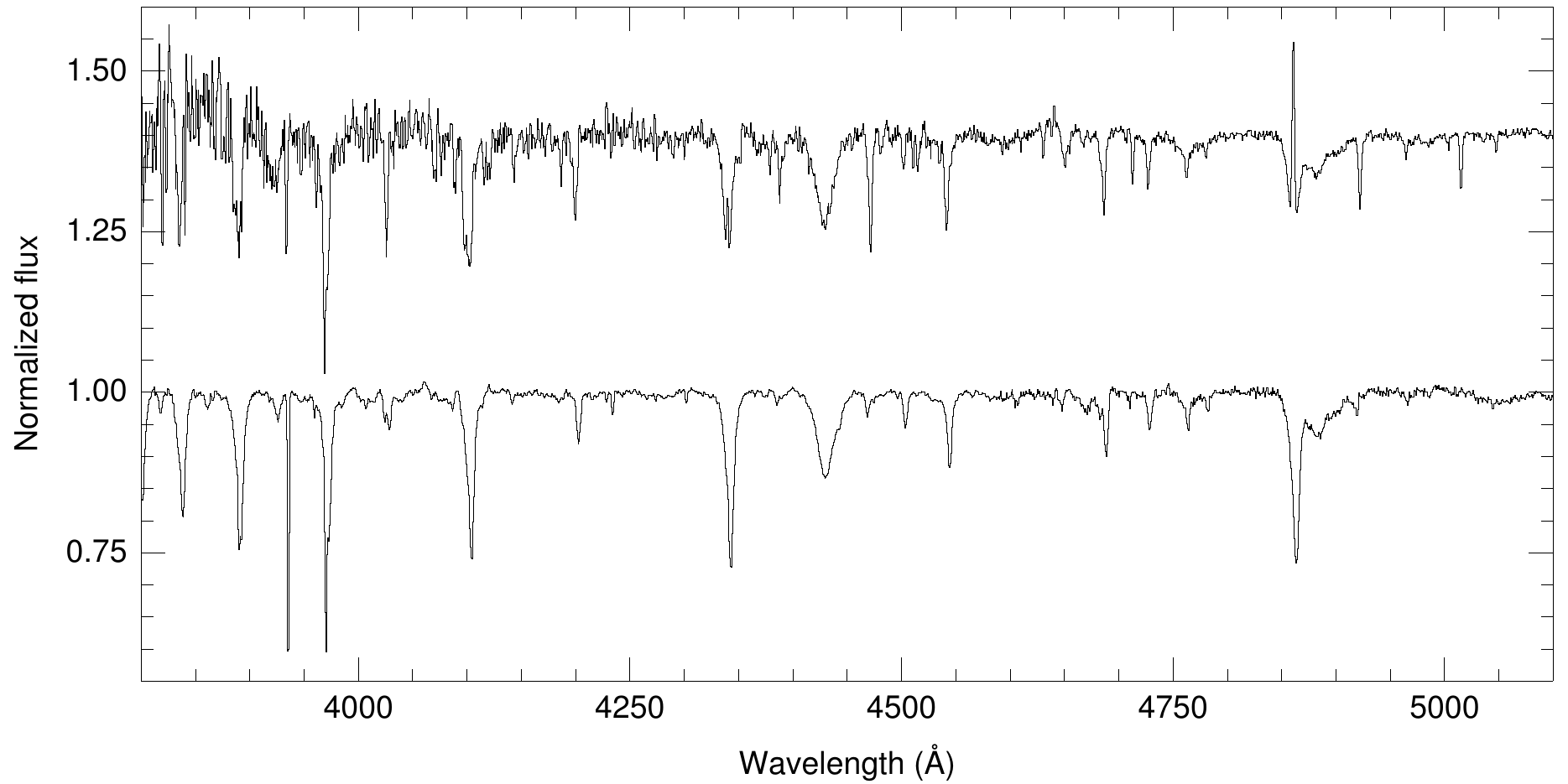}}
\caption{Two examples of GOSSS GTC spectra. [top] A newly discovered O7.5~III((f))p star, whose spectrum likely reveals a strong magnetic 
field, a rarity among O stars. [bottom] A new O3~V((f*)) + O9.5~V: spectroscopic binary system. The primary is the earliest known O dwarf in the 
northern hemisphere. }
\label{figure1}
\end{figure}

\begin{figure}
\centerline{\includegraphics*[height=7.2cm, bb=28 -70 566 566]{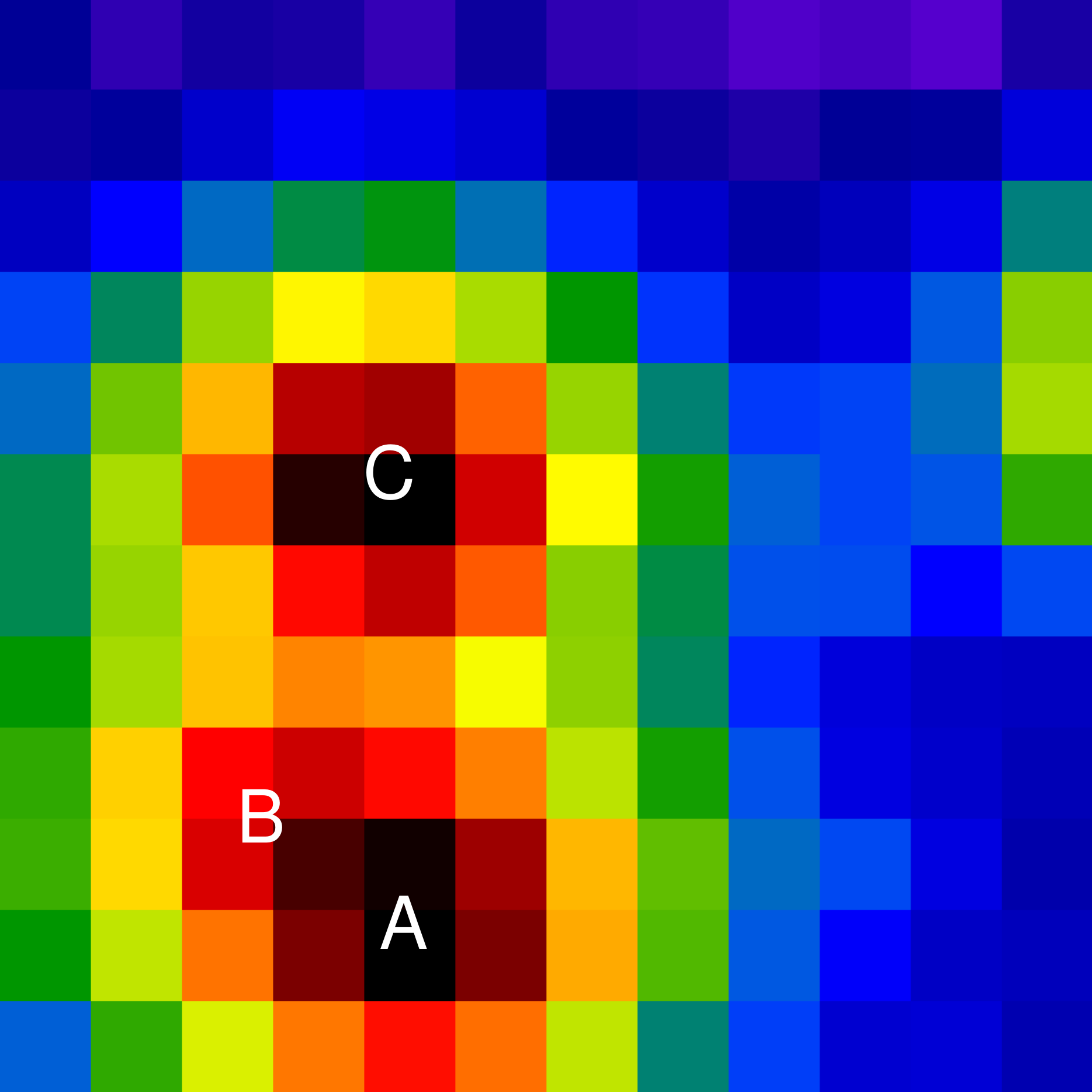} \
            \includegraphics*[height=7.0cm]{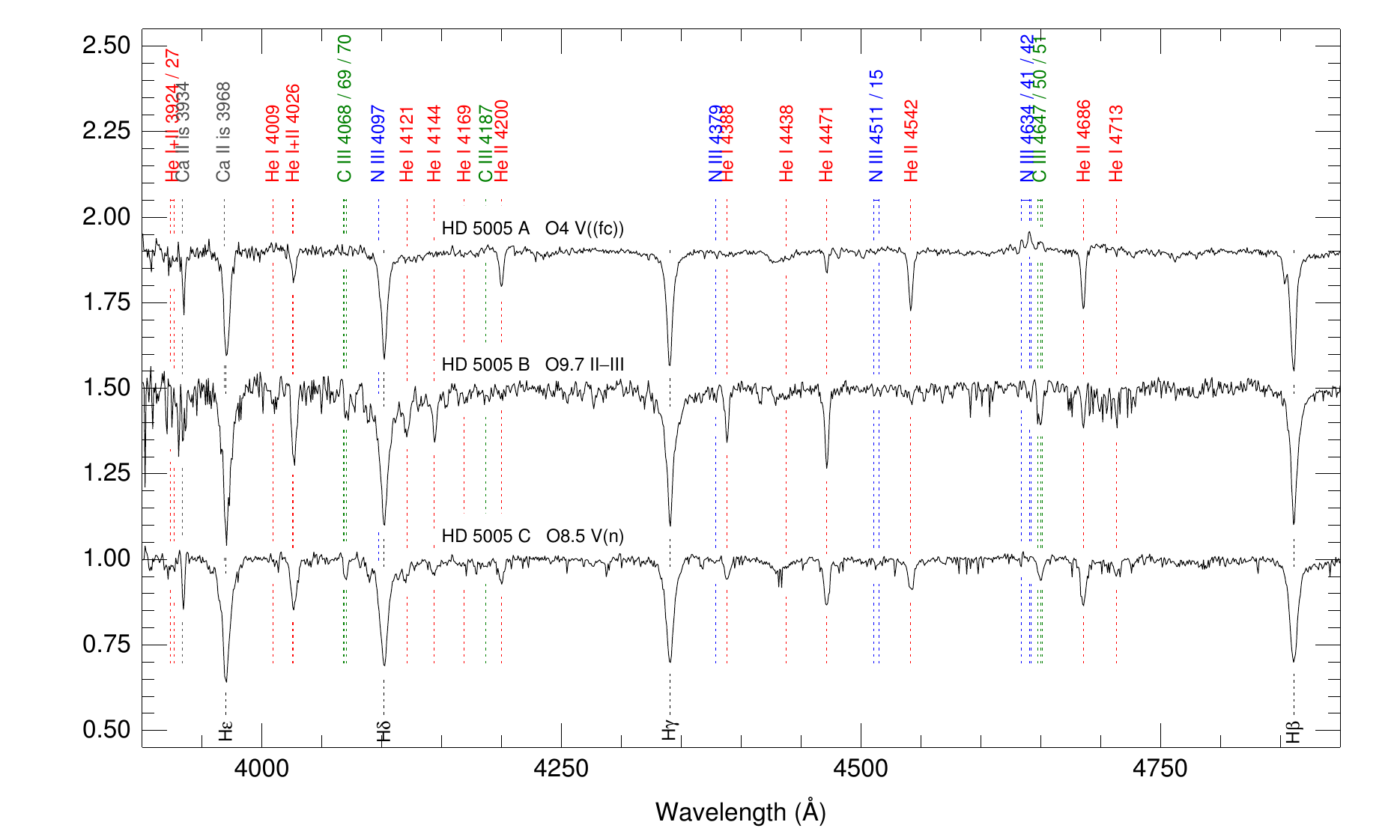}}
\caption{[left] Wavelength-collapsed FRODOspec@Liverpool image of the multiple system HD~5005~A+B+C (a D component
is located just outside the field to the right). The letters indicate the positions of the three stars. The field size is 
9\farcs84$\times$9\farcs84 and corresponds to a grid of 12$\times$12 fibers. [right] Pipeline-extracted spectra of the three stars.
The pipeline fits a triple Moffat profile at each wavelength point and iterates after fitting a polynomial in wavelength to 
the PSF parameters.}
\label{figure2}
\end{figure}

\section{GOSSS-III}

$\,\!$\indent We have recently published the third major block of the GOSSS project 
(\href{http://adsabs.harvard.edu/abs/2016ApJS..224....4M}{Ma{\'\i}z Apell\'aniz et al. 2016}). It includes the second of the
project data releases (GOSSS DR2.0).
The paper presents 142 additional stellar systems with O stars, for a total of 590 in GOSSS DR2.0.
Among them, there are 20 new O stars and 11 new SB2 systems: 6 of O+O type and 5 of O+B type. GOSSS-III also shows
revisions of previous GOSSS spectral types, including adaptations to the OV/OVz scheme
of \href{http://adsabs.harvard.edu/abs/2016AJ....152...31A}{Arias et al. (2016)}.
We also present egregious errors in the literature i.e. late-type stars classified as O type and
introduce luminosity class IV for spectral types O4-O5.5.

\section{MGB}

$\,\!$\indent MGB is a code that attacks spectral classification: 
\href{http://adsabs.harvard.edu/abs/2012ASPC..465..484M}{Ma{\'\i}z Apell\'aniz et al. (2012)}. It can be used to apply classical visual 
(non-automatic) spectral classification by interactively comparing with a standard grid. Four parameters can be adjusted:

\begin{itemize}
 \item Spectral subtype (horizontal classification).
 \item Luminosity class (vertical classification).
 \item n index (broadening).
 \item Alternative standards at each grid point (e.g. ONC or f variants).
\end{itemize}

\begin{figure}
\centerline{\includegraphics*[width=1.0\linewidth]{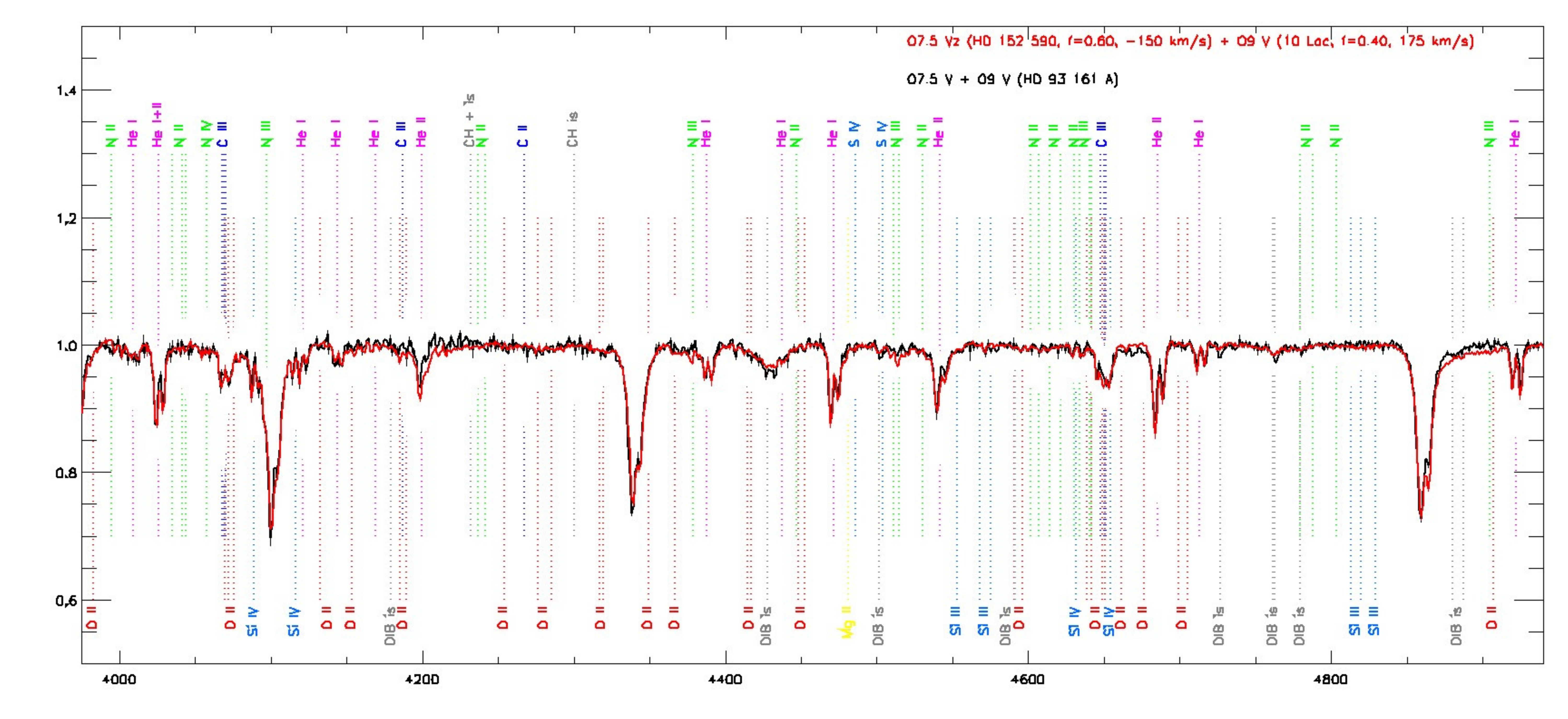}}
\caption{Example of fitting an SB2 system with MGB. Eight parameters can be adjusted: the spectral subtypes, 
luminosity classes, and velocities of both the primary and secondary, the flux fraction of the secondary, and 
the rotation index n. Here HD~93\,161~A (black) is fitted with a combination (red) of 60\% of HD~152\,590 and 
40\% of 10~Lac separated by 325~km/s.}
\label{figure3}
\end{figure}

The code includes fitting of SB2 systems (Figure~\ref{figure3}).

Here we introduce a new version of the code, MGB v2.0, which is available now from \url{http://jmaiz.iaa.es}.
The latest version includes a new default standard grid, OB2500 v3.0, which covers the O2-O9.7 spectral subtypes with GOSSS data (see below).
Other grids (O-type or other) from different on-going high-resolution surveys (e.g. IACOB, OWN, IACOBsweG) or from atmosphere models
can also be used.

\section{The GOSSS standard grid}
 
$\,\!$\indent We present a new version of the GOSSS standard grid, OB2500 v3.0, which is integrated with MGB (see above).
It covers the spectral subtypes from O2 to O9.7 and the luminosity classes from Vz to Ia (Table~2).
There are two types of gaps in the grid: non-existing types (blank) and standards not yet found (\ldots).
It is similar to OB2500 v2.0, the grid in \href{http://adsabs.harvard.edu/abs/2015hsa8.conf..603M}{Ma{\'{\i}}z Apell{\'a}niz (2015)}, 
but with small changes introduced by \href{http://adsabs.harvard.edu/abs/2016ApJS..224....4M}{Ma{\'{\i}}z Apell{\'a}niz et al. (2016)}:
addition of the Vz luminosity class, definition of luminosity class IV for O4-O5.5, and introduction of new standards.
It is available from \url{http://jmaiz.iaa.es} with MGB v2.0.
We plan to have a future extension to A0 (including all B stars) and luminosity class Ia+.

\section{The GOSSS future}

$\,\!$\indent Our plans for the future include:

\begin{enumerate}
 \item We will extend GOSSS from 590 O stars to $\sim$1500 in the next decade. 
 \item We will publish another 3000 spectra and spectral types (mostly B stars) in the same period.
 \item We expect to discover several tens of new SB2 systems.
 \item We will combine our results with Gaia and other surveys to derive the 6-D distribution of O stars in the solar neighborhood.
 \item We will use photometric surveys (e.g. GALANTE) to expand the sample.
 \item A complete sample is expected for the two outer Galactic quadrants (low extinction) at the end of survey. The two inner quadrants 
       require (a) multifiber surveys such as WEAVE and (b) IR surveys.
 \item We will study the spatial distribution of dust as a function of grain size.
 \item We will produce standard grids for the whole OB spectral range.
\end{enumerate}

\section*{References}

\begin{itemize}
 \item \href{http://adsabs.harvard.edu/abs/2016AJ....152...31A}{Arias, J. I. et al. 2016, {\it AJ} {\bf 152}, 31}. 
 \item \href{http://adsabs.harvard.edu/abs/2004ApJS..151..103M}{Ma{\'{\i}}z Apell{\'a}niz, J. et al. 2004, {\it ApJS} {\bf 151}, 103}. 
 \item \href{http://adsabs.harvard.edu/abs/2011hsa6.conf..467M}{Ma{\'{\i}}z Apell{\'a}niz, J. et al. 2011, {\it Highlights of Spanish Astrophysics VI}, 467}. 
 \item \href{http://adsabs.harvard.edu/abs/2012ASPC..465..484M}{Ma{\'{\i}}z Apell{\'a}niz, J. et al. 2012, {\it ASPC} {\bf 465}, 484}. 
 \item \href{http://adsabs.harvard.edu/abs/2015hsa8.conf..603M}{Ma{\'{\i}}z Apell{\'a}niz, J. et al. 2015, {\it Highlights of Spanish Astrophysics VIII}, 603}. 
 \item \href{http://adsabs.harvard.edu/abs/2016ApJS..224....4M}{Ma{\'{\i}}z Apell{\'a}niz, J. et al. 2016, {\it ApJS} {\bf 224}, 4 (GOSSS-III)}. 
 \item \href{http://adsabs.harvard.edu/abs/2008RMxAC..33...56S}{Sota, A. et al. 2008, {\it RevMexAA (SC)} {\bf 33}, 56}. 
 \item \href{http://adsabs.harvard.edu/abs/2011ApJS..193...24S}{Sota, A. et al. 2011, {\it ApJS} {\bf 193}, 24 (GOSSS-I)}. 
 \item \href{http://adsabs.harvard.edu/abs/2014ApJS..211...10S}{Sota, A. et al. 2014, {\it ApJS} {\bf 211}, 10 (GOSSS-II)}. 
\end{itemize}

\vfill

\eject

\addtolength{\textheight}{3cm}
\addtolength{\textwidth}{4cm}
\addtolength{\oddsidemargin}{-1.5cm}
\addtolength{\evensidemargin}{-1.5cm}

\begin{landscape}
Table 2. The OB2500 v3.0 grid of standards. \\

\footnotesize
\begin{tabular}{lllllllll}
\\
\hline
 & \multicolumn{1}{c}{Vz} & \multicolumn{1}{c}{V} & \multicolumn{1}{c}{IV} & \multicolumn{1}{c}{III} & \multicolumn{1}{c}{II} & \multicolumn{1}{c}{Ib} & \multicolumn{1}{c}{Iab/I} & \multicolumn{1}{c}{Ia} \\
\hline
O2   &                    &                       &                        &                          &                         &                   & {\it HD 93\,129 AaAb} &                    \\
\hline
O3   & {\it HD 64\,568}   & \nodata               &                        & \nodata                  &                         &                   &      Cyg OB2-7        &                    \\
\hline
O3.5 & {\it HD 93\,128}   & \nodata               &                        & {\it Pismis 24-17}       &                         &                   & {\it NGC 3603 HST-48} &                    \\
\hline
O4   & {\it HD 96\,715}   & {\bf HD 46\,223}      & {\bf HD 168\,076 AB}   & \nodata                  &                         &                   &      HD 15\,570       &                    \\
     &                    &                       & {\it HD 93\,250 AB}    &                          &                         &                   &      HD 16\,691       &                    \\
     &                    &                       &                        &                          &                         &                   &      HD 190\,429 A    &                    \\
\hline
O4.5 & \nodata            &      HD 15\,629       &      HD 193\,682       & \nodata                  &                         &                   &      HD 14\,947       &                    \\
     &                    & {\it HDE 303\,308 AB} &                        &                          &                         &                   &      Cyg OB2-9        &                    \\
\hline
O5   & {\bf HD 46\,150}   & {\it HDE 319\,699}    & {\bf HD 168\,112 AB}   & {\it HD 93\,843}         &                         &                   & {\it CPD -47 2963 AB} &                    \\
\hline
O5.5 & \nodata            & {\it HD 93\,204}      & \nodata                & \nodata                  &                         &                   &      Cyg OB2-11       &                    \\
     &                    &                       &                        &                          &                         &                   & {\it ALS 18\,747}     &                    \\
\hline
O6   &      HD 42\,088    & {\bf ALS 4880}        & {\it HD 101\,190 AaAb} &      HDE 338\,931        &      HDE 229\,196       & \nodata           & \nodata               & {\bf HD 169\,582}  \\
     & {\it HDE 303\,311} & {\it CPD -59 2600}    &                        &                          &                         &                   &                       &                    \\
\hline
O6.5 & {\it HD 91\,572}   & {\bf HD 167\,633}     & {\it HDE 322\,417}     &      HD 190\,864         & {\bf HD 157\,857}       & \nodata           & \nodata               & {\it HD 163\,758}  \\
     &                    &      HD 12\,993       &                        & {\it HD 96\,946}         &                         &                   &                       &                    \\
     &                    &                       &                        & {\it HD 152\,723 AaAb}   &                         &                   &                       &                    \\
     &                    &                       &                        & {\it HD 156\,738 AB}     &                         &                   &                       &                    \\
\hline
O7   & {\it HD 97\,966}   & {\it HD 93\,146 A}    &      ALS 12\,320       &      Cyg OB2-4 A         & {\it HD 94\,963}        & {\it HD 69\,464}  & \nodata               & \nodata            \\
     & {\it CPD -58 2620} & {\it HD 93\,222 AB}   &                        & {\it HD 93\,160 AB}      & {\it HD 151\,515}       &      HD 193\,514  &                       &                    \\
     &      HDE 242\,926  &                       &                        &                          &                         &                   &                       &                    \\
     & {\it HD 91\,824}   &                       &                        &                          &                         &                   &                       &                    \\
\hline
O7.5 & {\it HD 152\,590}  &      HD 35\,619       & {\it HD 97\,319}       & {\it HD 163\,800}        &      HD 34\,656         &      HD 17\,603   &      HD 192\,639      & \nodata            \\
     &                    &                       &                        &                          & {\bf HD 171\,589}       & {\it HD 156\,154} & {\bf 9 Sge}           &                    \\
\hline
O8   & {\it HDE 305\,539} & {\it HD 101\,223}     & {\it HD 94\,024}       & {\it HDE 319\,702}       & {\it 63 Oph}            & {\bf BD -11 4586} &      HD 225\,160      & {\it HD 151\,804}  \\
     & {\it HDE 305\,438} & {\it HD 97\,848}      & {\it HD 135\,591}      & {\bf $\lambda$ Ori A}    &                         &                   &                       &                    \\
     &                    &      HD 191\,978      &                        &                          &                         &                   &                       &                    \\
\hline
O8.5 &                    & {\it HDE 298\,429}    & {\bf HD 46\,966 AaAb}  & {\it HD 114\,737 AB}     & {\it HD 75\,211}        & {\it HD 125\,241} & \nodata               & {\it HDE 303\,492} \\
     &                    &      HD 14\,633 AaAb  &                        &      HD 218\,195 A       &      HD 207\,198        &                   &                       &                    \\
     &                    & {\bf HD 46\,149}      &                        &                          &                         &                   &                       &                    \\
     &                    & {\it HD 57\,236}      &                        &                          &                         &                   &                       &                    \\
     &                    & {\it Trumpler 14-9}   &                        &                          &                         &                   &                       &                    \\
\hline
O9   &                    &      10 Lac           & {\it HD 93\,028}       & {\it HD 93\,249 A}       & {\it HD 71\,304}        &      19 Cep       &      HD 202\,124      &      $\alpha$ Cam  \\
     &                    &      HD 216\,898      & {\it CPD -41 7733}     &      HD 24\,431          & {\it $\tau$ CMa AaAb}   &                   & {\it HD 152\,249}     &                    \\
     &                    & {\it CPD -59 2551}    &                        &                          &                         &                   &      HD 210\,809      &                    \\
\hline
O9.2 &                    & {\bf HD 46\,202}      & {\it HD 96\,622}       & {\it CPD -35 2105 AaAbB} &      ALS 11\,761        & {\it HD 76\,968}  & {\it HD 154\,368}     & {\it HD 152\,424}  \\
     &                    &      HD 12\,323       &                        &      HD 16\,832          &                         &                   & {\it HD 123\,008}     &                    \\
     &                    &                       &                        &                          &                         &                   &      HD 218\,915      &                    \\
\hline
O9.5 &                    &      AE Aur           &      HD 192\,001       & {\it HD 96\,264}         & {\bf $\delta$ Ori AaAb} & \nodata           &      HD 188\,209      & \nodata            \\
     &                    & {\it $\mu$ Col}       & {\it HD 93\,027}       &                          &                         &                   &                       &                    \\
     &                    &                       & {\it HD 155\,889 AB}   &                          &                         &                   &                       &                    \\
\hline
O9.7 &                    & {\bf $\upsilon$ Ori}  &      HD 207\,538       &      HD 189\,957         & {\it HD 68\,450}        & {\bf HD 47\,432}  &      HD 225\,146      &      HD 195\,592   \\
     &                    &                       &                        & {\it HD 154\,643}        & {\it HD 152\,405}       & {\it HD 154\,811} & {\it $\mu$ Nor}       & {\it GS Mus}       \\
     &                    &                       &                        &                          &      HD 10\,125         & {\it HD 152\,147} & {\it HD 104\,565}     &                    \\
     &                    &                       &                        &                          &                         &                   &      HD 191\,781      &                    \\
\hline
Notes & \multicolumn{8}{l}{Normal, {\it italic}, and {\bf bold} typefaces are used for stars with $\delta > +20\degr$, $\delta < -20\degr$, and the equatorial intermediate region, respectively.}
\end{tabular}
\end{landscape}

\end{document}